\documentclass[compsoc,journal]{mytran}

\RequirePackage[compress]{cite}
\usepackage[utf8]{inputenc}
\usepackage[T1]{fontenc}
\usepackage[english]{babel}
\usepackage{textcomp}
\usepackage{url}
\usepackage{graphicx}
\usepackage{float}
\usepackage{multirow}
\usepackage{verbatim}

\usepackage{multicol}
\usepackage{xcolor}

\usepackage{amsmath, amsfonts, mathtools, amsthm,amssymb,accents}

\usepackage{subfig}

\newcommand{\mQ}{\mathcal Q}

\newcommand{\citep}{\cite}
\newcommand{\citet}{\cite}

\usepackage{animate}

\usepackage{color,array,verbatim,algpseudocode,bm,algorithm}
\definecolor{menucolor}{rgb}{0.1,0.52,0.47}
\definecolor{urlcolor}{rgb}{0.85,0.37,0.01}
\definecolor{runcolor}{rgb}{0.46,0.44,0.701}
\definecolor{filecolor}{rgb}{0.2,0.5,0.01}
\definecolor{linkcolor}{rgb}{0.12,0.47,0.70}
\definecolor{citecolor}{rgb}{0.55,0.36,0.01}
\definecolor{anchorcolor}{rgb}{0.4,0.4,0.4}
\usepackage[colorlinks=true, urlcolor=urlcolor, linkcolor=linkcolor,citecolor=citecolor,filecolor=filecolor, anchorcolor=anchorcolor,menucolor=menucolor]{hyperref}
\usepackage{nameref}
\usepackage{zref-xr}

\makeatletter
\newcommand\VeryLarge{\@setfontsize\Huge{16}{16}}
\newcommand\smaller{\@setfontsize\small{5}{5}}
\makeatother   

\def\subparagraph{} 
\usepackage{titlesec}
\titlespacing*{\section}{0pt}{*1}{*1}
\titlespacing{\subsection}{0pt}{*1}{*1}
\titlespacing{\subsubsection}{0pt}{*1}{*1}

\titleformat{\paragraph}[runin]{\normalfont\bfseries}{\theparagraph}{0.5em}{}[:]

\IEEEaftertitletext{\vspace{-3\baselineskip}} 



\renewcommand{\baselinestretch}{0.9}
\normalsize

\begin{document}
\title{
  \vspace{-0.3in}
  \VeryLarge Boxplots and quartile plots for grouped and periodic angular data}
\author{
  \vspace{-0.15in}
  Joshua D. Berlinski, Fan Dai and Ranjan Maitra\\
  \vspace{-0.45in}
  \thanks{J. D. Berlinski and R. Maitra are with the Department of Statistics, Iowa State University, Ames, Iowa 50011, USA, and F. Dai is with the Department of Mathematical Sciences, Michigan Technological University, Houghton, MI 49931, USA. Corresponding author email: {\em maitra@iastate.edu}}
    
\thanks{{\bf Abbreviations:} 1D, one-dimensional; 2D, two-dimensional; 3D, three-dimensional; IQR, interquartile range.}
}
\maketitle
\begin{abstract}
Angular observations, or observations lying on the unit circle, arise in many disciplines and require special care in their description, analysis, interpretation and visualization. We provide methods to construct concentric circular boxplot displays of 
distributions of groups of angular data. The use of concentric boxplots brings challenges of visual perception, so we set the boxwidths to be inversely proportional to the square root of their distance from the centre. A perception survey supports this scaled boxwidth choice. For a large number of groups, we propose  circular quartile plots. A three-dimensional toroidal display is also implemented for periodic  angular distributions. We illustrate our methods on datasets in (1) psychology, to display motor resonance under different  conditions, (2) genomics, to understand the distribution of peak  phases for ancillary clock genes, and  (3) meteorology and wind turbine  power generation, to study the changing and periodic distribution of wind direction over the course of a year.

\end{abstract}
\vspace{-0.1in}

\section{Introduction}
Angular data, also called circular or directional data, have observations that are measured as angles \citep{fisher1993} and arise in fields~\citep{pewsey18} such as biology \citep{deota2023}, psychology \citep{cremers2018}, meteorology \citep{klink1999, koch2004},  political science~\citep{gillandhangartner10}, digital imaging~\citep{berlinskiandmaitra25}, and so on. 
The angle's periodicity means that care must be taken when analyzing directional  data. 
For example, the mean of $1^\circ$ and $359^\circ$ calculated using linear statistics is exactly antipodal to what is intuitive and accounts  for the directionality of the data.
This dichotomy is important when designing visualizations for angular data.
For example, Figure 1 of \citet{gillandhangartner10}  illustrates hourly data,
collected over 24 hours, on gun-related crimes over 1987-88 in Pittsburgh,
Pennsylvania, and displays how representing the data using a linear histogram shows
events to start at a high level at midnight, then decrease and finally return to
a high level as the day ends. It misses the connection between the left and
the right end-points and obscures the fact that gun crimes  generally peaked at  between 10 pm and 2 am in 1988. On the other hand, displaying the data through a rose diagram preserves the upturn around midnight and can better inform legislative policy and
law-enforcement priorities.

Rose diagrams are histogram-like displays useful for displaying
distributions of angles \citep{sundari2020} 
but are limited only to a single group or population. Comparisons  between populations can be done only side-by-side, but such an approach  hinders accurate comparison of the features of angular data. Figure 1 of \citet{cremers2018} plots individual points on a circular axis, denoting the direction in which subjects walk after being
instructed to move east. While succinct and simple, such a display (essentially, a directional analog of a dotplot) is not a summary, besides being impractical for larger datasets. Figure 2 of \citet{buttarazzi2018} plots similar angular observations atop each other outwardly from the center of the circle, and compares this approach to rose diagrams, smoothed rose diagrams, and circular boxplots. 
While these tools display the angular nature of circular data, it is not always easy to quickly identify important features of the data.

In linear statistics, a boxplot \citep{tukey1977} effectively displays the main features   of a dataset, notably its centre, spread, skewness, and outliers. 
For angular data, we must account for the wrapping while deriving and calculating circular analogs of the ingredients of a boxplot.
Incorporating this wrapping can be challenging, so some practitioners simply rotate all
observations by a constant value such that the $0^\circ$ direction is not included in the range of the rotated data\citep{buttarazzi2018}. They then create a boxplot using linear statistics, visually changing axis labels to account for
the rotation and optionally visually transforming the axes to be circular. This
approach fails when data span the entire circle, but also because the calculated quantities are mostly far from correct.  \citet{buttarazzi2018} filled this lacuna by introducing a circular ordering of observations from which circular summary statistics analogous to those used in Tukey's boxplot can be derived. 

The circular boxplot of \citet{buttarazzi2018} only displays  angular data from one population, so any comparison between multiple groups of data can only be done side-by-side. With many groups, such plots can be cumbersome to reference. Also, as seen shortly in Figure~\ref{fig1}, side-by-side plots can make it challenging to compare  datasets with only subtle differences. 

This article makes two sets of contributions to circular boxplot displays. The first set of contributions, in Section~\ref{sec:2d}, allows for multiple groups to be displayed on the same plot by constructing the circular boxplots concentrically. 
However, our extension of the visualization methodology of \cite{buttarazzi2018} is not without challenges, notably because the area of an arced box changes depending on
its distance from the centre of the circle. Our remedy here is to match the
visual perception of identical boxes at different distances from the centre. Indeed, a short survey on perception of grouped circular boxplots supports this choice. For larger numbers of groups, we adapt quartile plots~\cite{tufte1998} to the directional case. 
Our second contribution, in Section~\ref{sec:3d}, uses 3D visualization to display the distribution of angles over a period, such as for  displaying the distribution of wind direction at a location over the course of the year. Specifically, we introduce 3D toroidal display boxplots and quartile plots to address periodicity of the angular wind direction and time. This article concludes with some discussion. A R\citep{R} package {\bf CircularBoxplots} implements our methods. 

\section{Displaying grouped circular data}
\label{sec:2d}
We first discuss methods for displaying grouped angular data. We provide background,  our development and its use in describing a few datasets.
\subsection{Background and preliminaries}
\label{prelim}
\subsubsection{Linear boxplots}
\label{linear-boxplot}
A linear boxplot is constructed~\citep{tukey1977} by first ranking the observations $(X_1,X_2,\ldots,X_n)$ both in  ascending, and separately, also in descending order. Each observation is then assigned a {\em depth} defined as the minimum of the two ranks.
These depth values are used to identify observations in the dataset (with interpolation as needed) to form a five-number data summary, comprising the minimum $X_{(1)}$, first quartile $\mQ_1$, median $\mQ_2$, third quartile $\mQ_3$, and the maximum $X_{(n)}$. A boxplot is constructed by drawing a box spanning $\mQ_1$ and $\mQ_3$, called {\em hinges}, with $\mQ_2$ indicated by a line drawn within the box and parallel to the shorter edge. Lines, called {\em whiskers}, are drawn on either side of the box to the furthest
 point that is within some constant, which is typically taken to be 1.5 times the length of the box ({\em i.e.}, the interquartile range or {\em IQR}) of the respective hinge. (The usual value of 1.5 ensures that around 0.7\% of observations lie outside the reach of these whiskers, under normal distributional assumptions for the data.)
Observations beyond the range of the box and the whiskers are plotted individually 
as {\em outliers}.

\subsubsection{Circular boxplots}
\label{circ-boxplot}
Extending boxplots to angular data $\theta_1,\theta_2\ldots,\theta_n{\in}[0,2\pi)$ is not immediate because of the amorphous definition of the origin, and the wraparound effect.  So \cite{buttarazzi2018} formulated a notion of depth for angular data. The starting point is \citet{mardia1999}'s definition of the {\em circular median} $\tilde\theta$ and (its antipodal point) the {\em circular antimedian} $\underaccent{\tilde}{\theta}$, where the pair $\{\tilde\theta,\underaccent{\tilde}{\theta}\}$ is defined so that about half of the $\theta_1,\theta_2,\ldots,\theta_n$ are encountered as we go clockwise from
$\tilde\theta$ to $\underaccent{\tilde}\theta$, and the other half are encountered in the clockwise direction from $\underaccent{\tilde}\theta$ to $\tilde\theta$.
Each observation $\theta_i$ is then ranked based on both its clockwise and counter-clockwise ordering from $\underaccent{\tilde}\theta$, from where the {\em circular depth} of each observation is defined as the lowest of these two ranks.

With these depth values assigned, a circular five-number summary can be
computed. Specifically, for a dataset with $n$ circular observations, the {\em circular
depth} of the median is $(1 + n) / 2$, with depths admitted in halves.
The hinges representing the quartiles are then defined as having circular depth
$(1 + \lfloor(1 + n)/2\rfloor)/2$, where $\lfloor x \rfloor$ is the largest
integer no larger than $x$.
The box is drawn on the shortest arc joining the two hinges. The observations to which arcs or {\em whiskers} extending from the
hinges should be drawn are determined as the last observations inside the upper and lower inner {\em fences} that  are computed by scaling the length of the box by a factor, 
proposed by \citet{buttarazzi2018} to be according to the concentration level of angular data, measured by the mean resultant length, $R=\sqrt{(\sum_{i=1}^{n}\cos\theta_i)^2 + (\sum_{i=1}^{n}\sin\theta_i)^2}$. (Because of the wraparound effect, and unlike for the linear boxplot, the
whiskers of a circular boxplot may span the entire circle.) Angular observations outside the lower and upper fences are denoted as {\em
outliers}. Further, \citet{buttarazzi2018} also provided some theoretical properties of the circular boxplot.

\begin{figure}[!h]
  \vspace{-0.2in}
  \centering.
\subfloat[$\pi_1=0.1,\pi_2=0.9$]{\label{fig1-b}
\includegraphics[width=0.3\columnwidth{}]{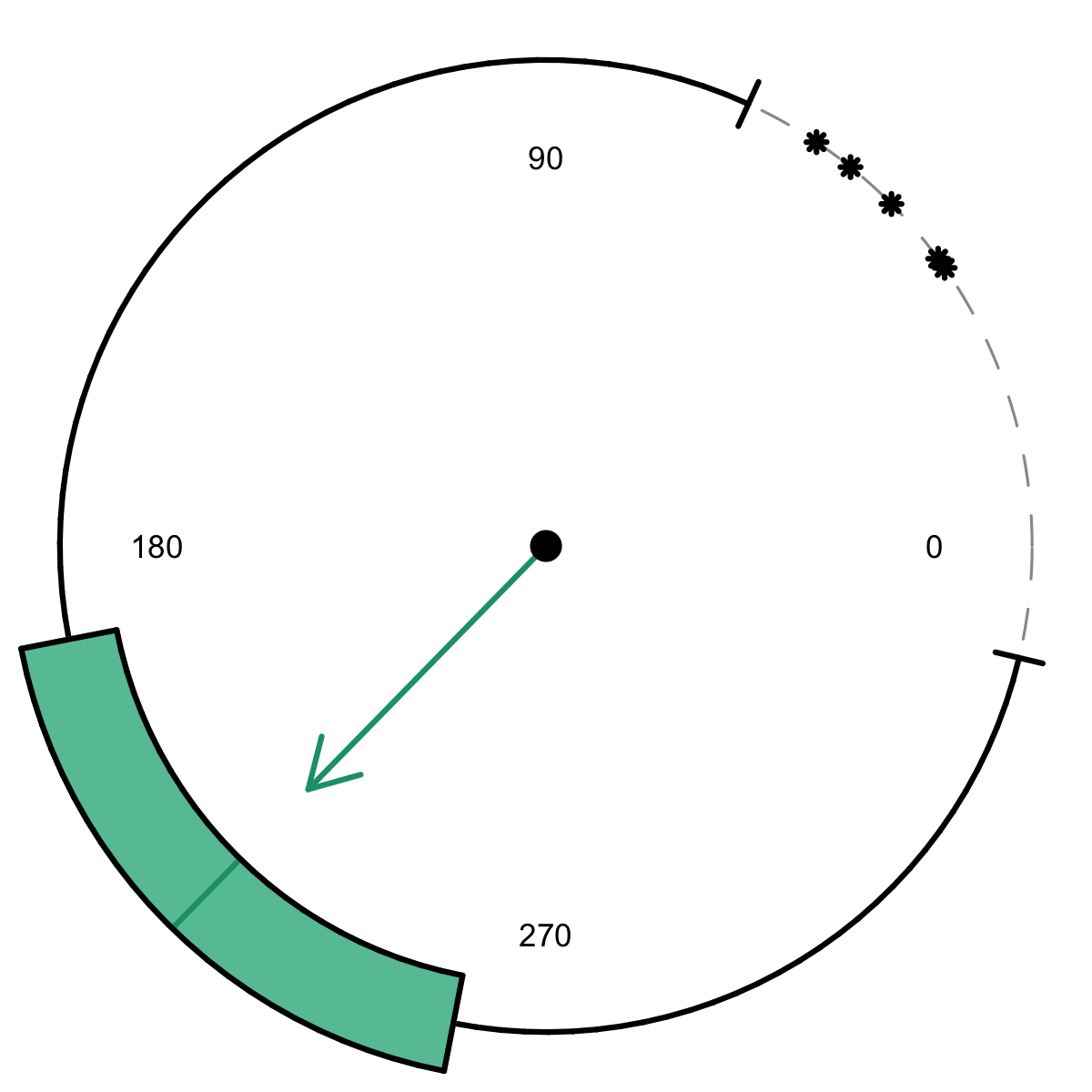}
}
\subfloat[$\pi_1=0.2,\pi_2=0.8$]{\label{fig1-c}
\includegraphics[width=0.3\columnwidth{}]{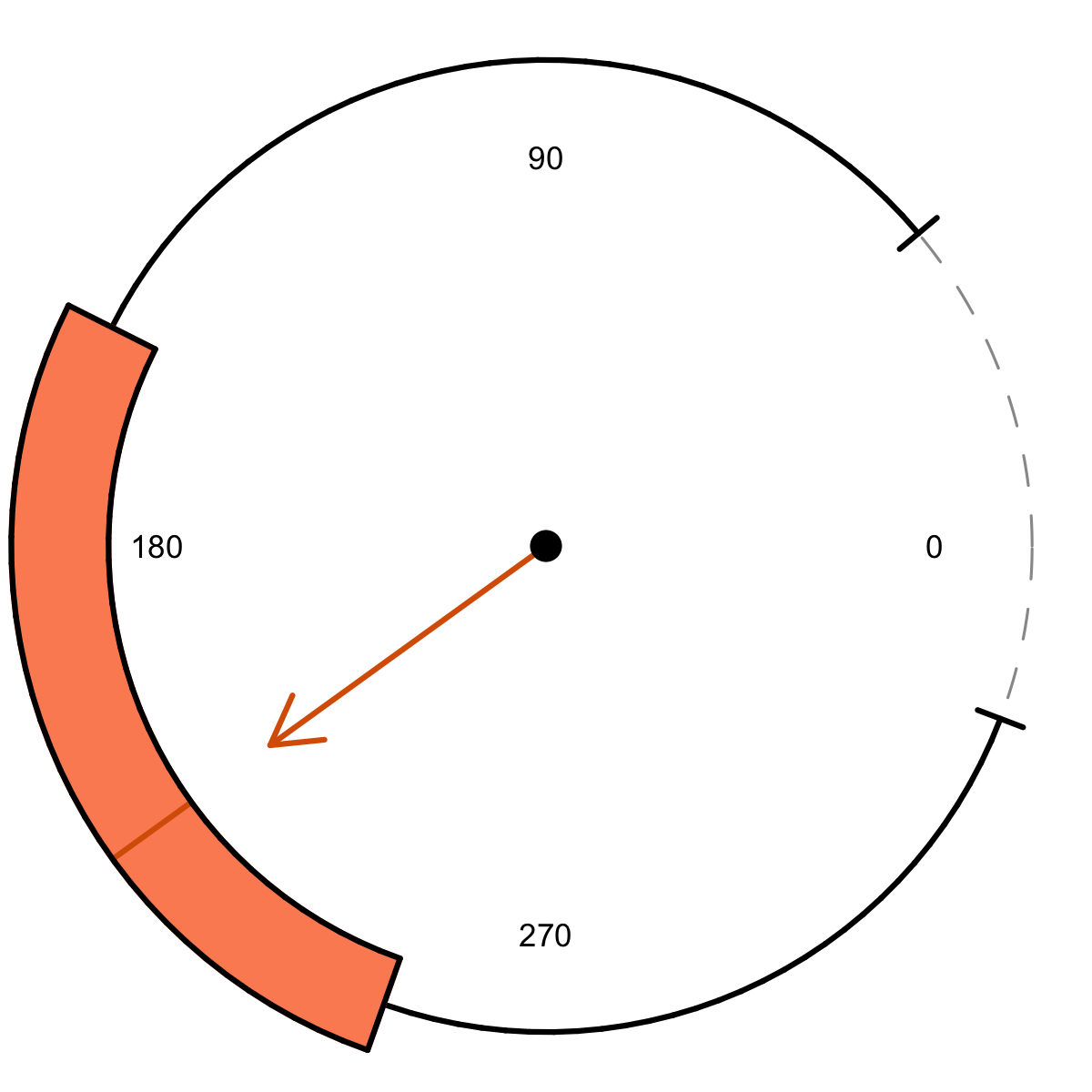}
}
\subfloat[$\pi_1=0.3,\pi_2=0.7$]{\label{fig1-a}
\includegraphics[width=0.3\columnwidth{}]{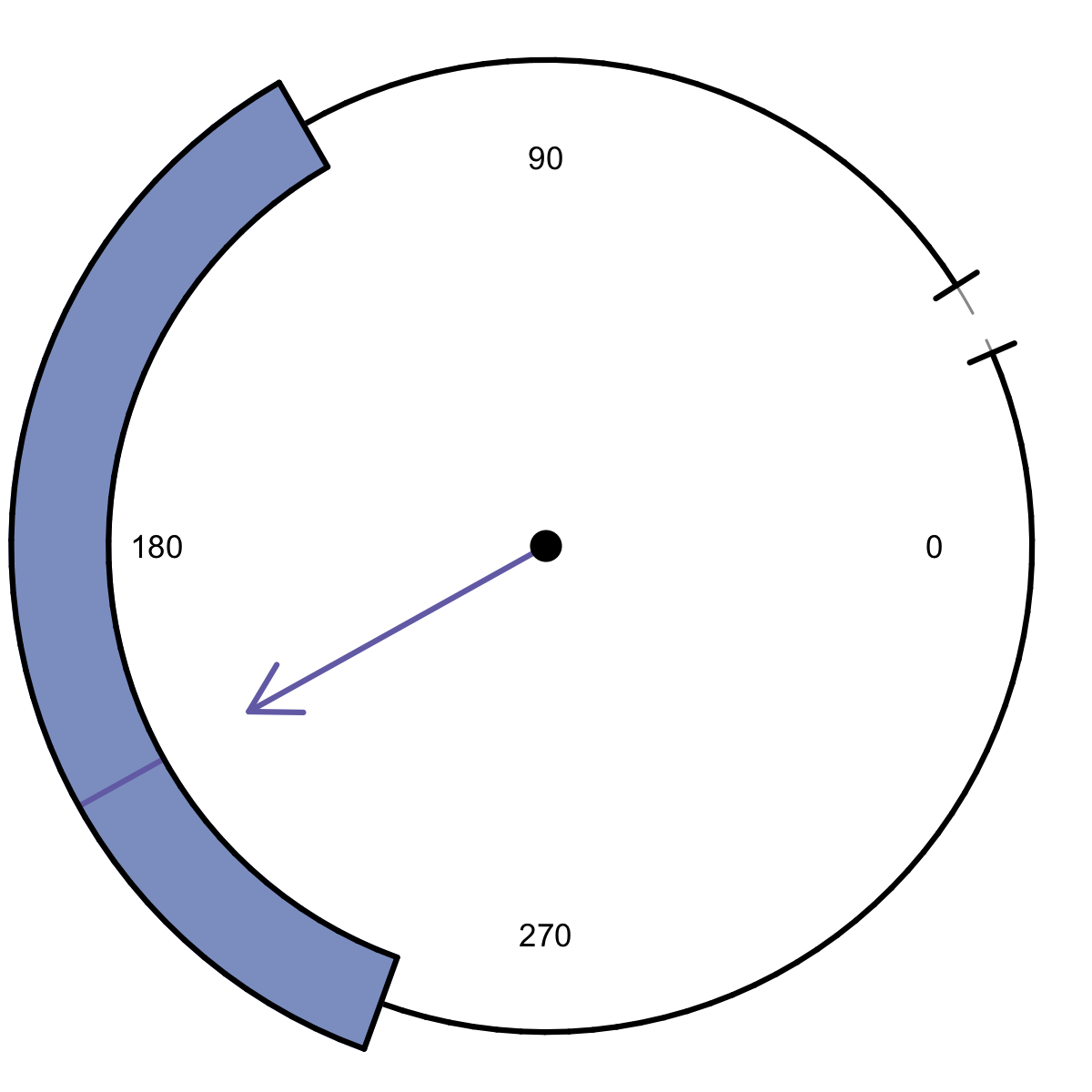}
}
\caption{Circular boxplots of samples from three different two-component mixture of  von Mises distributions. The mixing components have mean directions $95^\circ$ and $230^\circ$, and concentration parameters   $5.0$ and $2.5$, and mixing proportions $\pi_1$ and $\pi_2$, with values as specified.}
\label{fig1}
\end{figure}

\paragraph*{Limitations}
The implementation of \cite{buttarazzi2018} is only for one population, so comparisons between groups  of angular data require multiple circular boxplots presented side-by-side. Such a strategy is feasible only with a few groups from clearly dissimilar distributions, and even then it is not always possible to highlight minor distinctions. For consider Figure~\ref{fig1}, which displays circular boxplots of 
datasets simulated under three situations.
Each dataset is of 300 independent identically distributed pseudo-random
realizations  from a mixture of two von Mises
distributions~\cite{mardia1999}.
The only difference between the three samples is in the mixing proportions
($\pi_1,\pi_2$). 
Figure~\ref{fig1} shows that the distinctions between the distributions are  not easily appreciated from the side-by-side display. For instance, the distinctiveness of the medians in Figures \ref{fig1-b} and \ref{fig1-a} is unclear.  Also, while Figure \ref{fig1-c} is more-or-less symmetric, the relative circular skewnesses of the other two samples is unclear.

Both these questions could perhaps be  easily answered from Figure~\ref{fig1} by carefully estimating the relative sizes of the arced boxes in the boxplot. However, the main premise of any graphical display is that it should readily provide the practitioner with a good understanding and summary of the
important features of the data.
Therefore, our goal here is to provide such understanding and contrast between
grouped circular data by combining  circular boxplots from multiple groups into an integrated plot.
As we shall see, this comes with a few challenges that need to be addressed.

\subsection{ Grouped circular boxplots}
\label{group-circ-boxplot}
\begin{figure}[h]
    \vspace{-0.2in}
  \mbox{\subfloat[Same boxwidth]{\label{fig2-unscaled}\includegraphics[width=0.5\columnwidth]{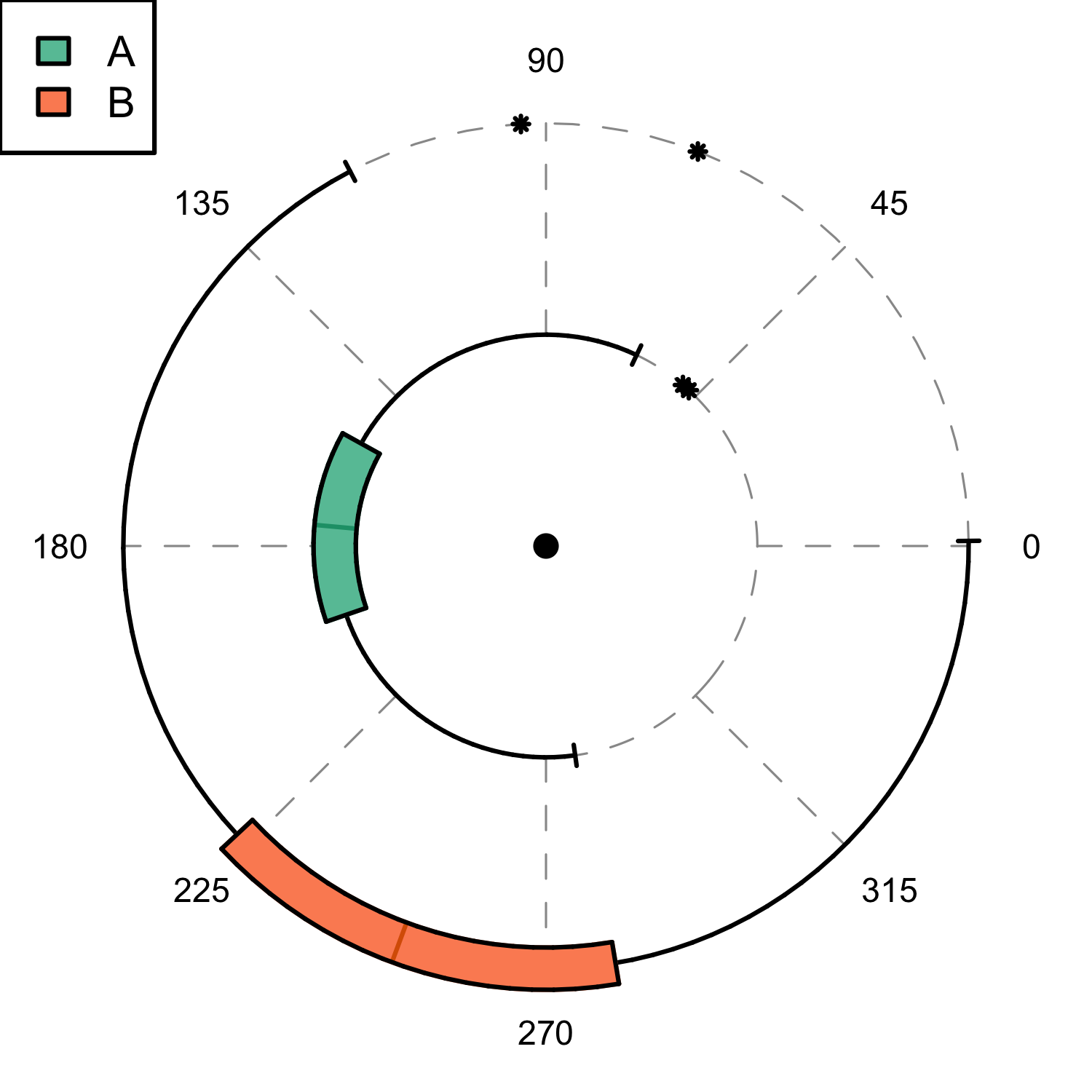}}%
  \subfloat[Scaled boxwidth]{\label{fig2-scaled}\includegraphics[width=0.5\columnwidth]{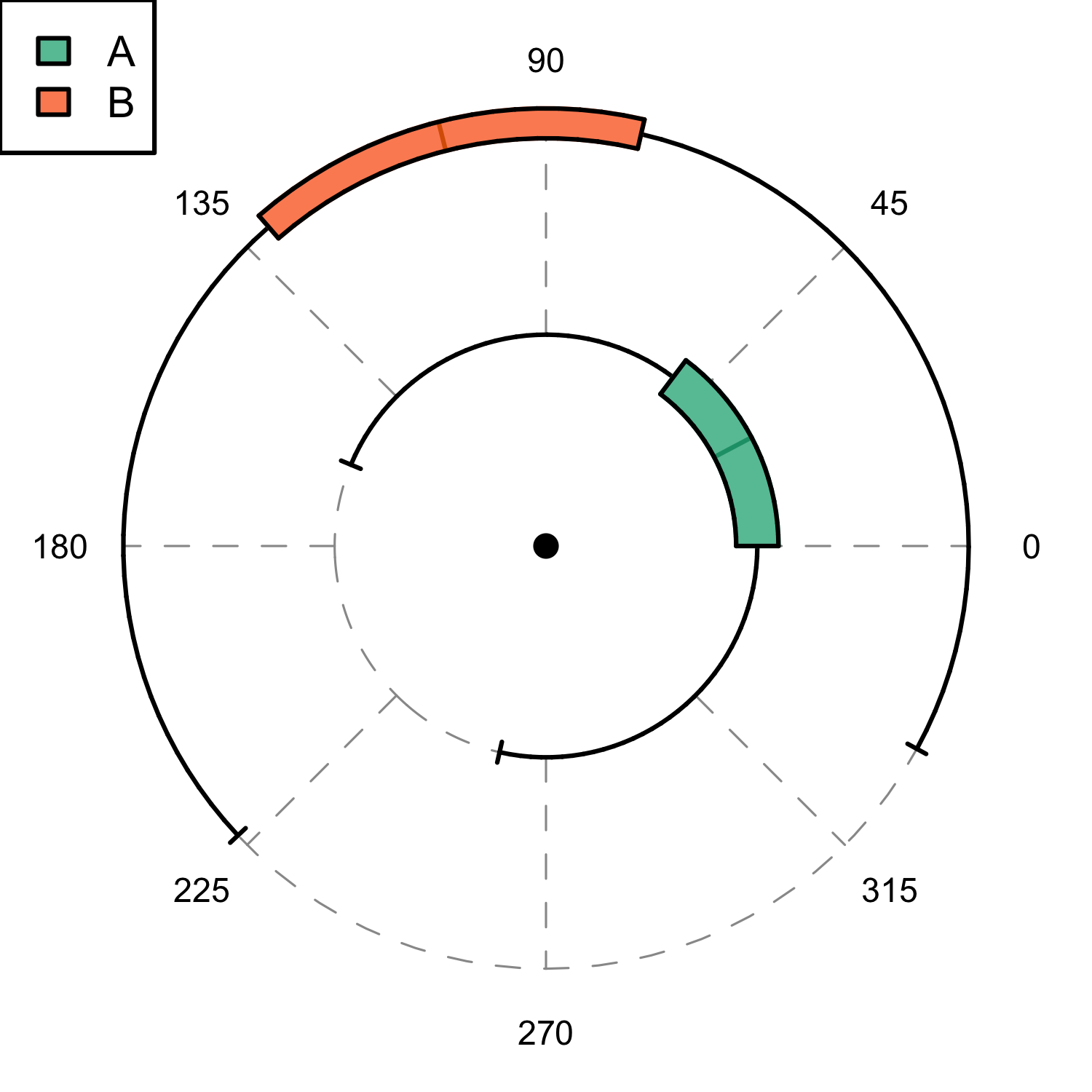}}}\\
  \subfloat[Results of the perception survey.]{\label{tab1}\small
  \begin{tabular}{c|ccc}
    \hline
    \multirow{2}{*}{\textbf{Scaled Width} (Figure \ref{fig2-scaled})} &
      \multicolumn{3}{c}{\textbf{Same Width} (Figure \ref{fig2-unscaled})}\\
      & $\delta_A > \delta_B$ & $\delta_B > \delta_A$ & $\delta_A\approx\delta_B$ \\
      \hline
    $\delta_A > \delta_B$ & 0 & 8 & 2 \\
    $\delta_B > \delta_A$  & 1 & 17 & 1\\
    $\delta_A\approx\delta_B$ & 2 & 24 & 9 \\
    \hline
  \end{tabular}
  }
  \caption{(a, b) Two grouped circular boxplots for two groups of angular observations (A for green and B for orange). In each plot, both groups have the same circular spread ($\delta_A\approx\delta_B$) but a different mean direction. The mean directions of (a) are rotated clockwise by $144^\circ$ in (b). Further, in (a), the two boxplots have the same boxwidth while in (b) the boxwidths are inversely proportional to the square root of their distance from the centre of the circle. (c) Results of an anonymous survey where respondents were asked to compare the circular spreads of the distributions of the two groups in each of the two plots.}
\label{fig2}
\end{figure}

The initial ingredients of a grouped circular boxplot are derived from  
Section~\ref{circ-boxplot} 
that is used to compute the circular summary statistics for each group.
The group-wise circular boxplots are then drawn concentrically.
For a small number of groups, an appropriately colored arrow pointing in the median
direction can be drawn per group, as in
\cite{buttarazzi2018}.
Figure~\ref{fig2-unscaled} presents an initial attempt at a grouped boxplot of two slightly circular-skewed groups of simulated circular data using this prescription. These two groups each had realizations from a two-component von Mises mixture distribution with concentration parameters $\kappa_1{=}7$ and $\kappa_2{=}3.5$ and mixing proportions $\pi_1{=}0.1$ and $\pi_2{=}0.9$. The mixture components of Group A had mean directions $\mu_1{=}95^\circ$ and $\mu_2{=}180^\circ$ while those of Group B had mean directions $\mu_1{=}170^\circ$ and $\mu_2{=}255^\circ$. 
The figure provides an indication of the perception challenges posed by the concentric arced boxes in the circular boxplots that is not present in the linear case. We see that perceptually the larger (by way of area) Group B box trains our visual cues to instinctively feel that the spread of the (outer) group is more than that of Group A when the two samples have the same circular {\em IQR} of $47^\circ$. 
And this, despite the helpful guides provided every $45^\circ$ apart. In the linear case, this is not an issue because boxplots placed next to each other and of the same boxwidth  (with height dictated by the scaled IQR) are easily compared.

For a possible solution, we adopt the fix that is traditionally employed in the context of histograms for linear data. Specifically, we note that the perception difference between the two data distributions in Figure \ref{fig2-unscaled} 
is because of the concentric circles, which result in the arc length of the box for each group depending on the radius at which it is drawn. When each boxwidth is the same, as in Figure \ref{fig2-unscaled}, the area of the box along the outer concentric circle is  more than that drawn along the inner one. Each box is a 2D representation of 1D data, and visual
perception is based on the area of the boxplot that may mislead (this is also same reason frequency histograms have their boxes plotted such that each box's area is proportional to that of the frequencies being compared). In Figure \ref{fig2-unscaled}, the outer box seems larger since it spans a longer total distance in the plot with the same total width. Since the box is meant to convey and subsequently compare the sizes of the inner
half of the dataset, boxes that cover the same arc should have the same area. Our solution  scales the boxwidth in each
boxplot to be inversely proportional to the radius along which it is drawn. 
We illustrate this approach in Figure \ref{fig2-scaled}, 
which displays 300 observations each from two groups from similar mixtures as before, with the only difference being that the means here are rotated by $144^\circ$ in the clockwise direction. The area of each box spanning the same arc is then essentially the same and  visually similar. So, it is more immediately evident that the two boxplots, and hence the distributions, are, but for the shift in their medians, the same.

We evaluated perception in using the same and the scaled boxwidths by 
posing Figures~\ref{fig2-unscaled} and
~\ref{fig2-scaled} in an anonymous online survey that asked respondents to separately  compare the circular dispersion of the two groups in each of the two boxplots. The survey was distributed to a target audience that chosen to have at least an introductory course in statistics, thereby ensuring exposure and familiarity with the (linear) boxplot. 
There were 64 responses, with results presented in Figure~\ref{tab1}. 
With regard to Figure~\ref{fig2-unscaled}, 49 (76.56\% of) respondents thought that Group B had a wider dispersion while only 12 (18.75\%) correctly perceived the dispersions of both groups to be indistinguishable. However, when looking at Figure~\ref{fig2-scaled}, 35 (54.7\%) of these respondents thought that 
the Groups A and B had the same circular spread while 19 (29.7\%) of them 
placed the spread of Group B as more than that of Group A. An exact one-sided McNemar's test \citet{McNemar_1947} that tested whether the scaling in Figure~\ref{fig2-scaled} 
substantially improved perception over the unscaled boxwidths in Figure~\ref{fig2-unscaled} was highly significant, reporting a $p$-value of $7.618\times10^{-6}$. A $95\%$ Monte Carlo confidence interval of this $p$-value is $(1.863\times10^{-9}, 0.0036)$. 
Our approach significantly improved perception, but almost 29.7\% of respondents saw Group B as having more circular spread than Group A, indicating some lingering perception challenges in  displaying grouped circular boxplots, 
But even then, the survey conclusively demonstrated the benefits of our boxwidth-scaling strategy for grouped circular boxplots.
\begin{figure}[!ht]
\centering  \includegraphics[width=0.65\columnwidth]{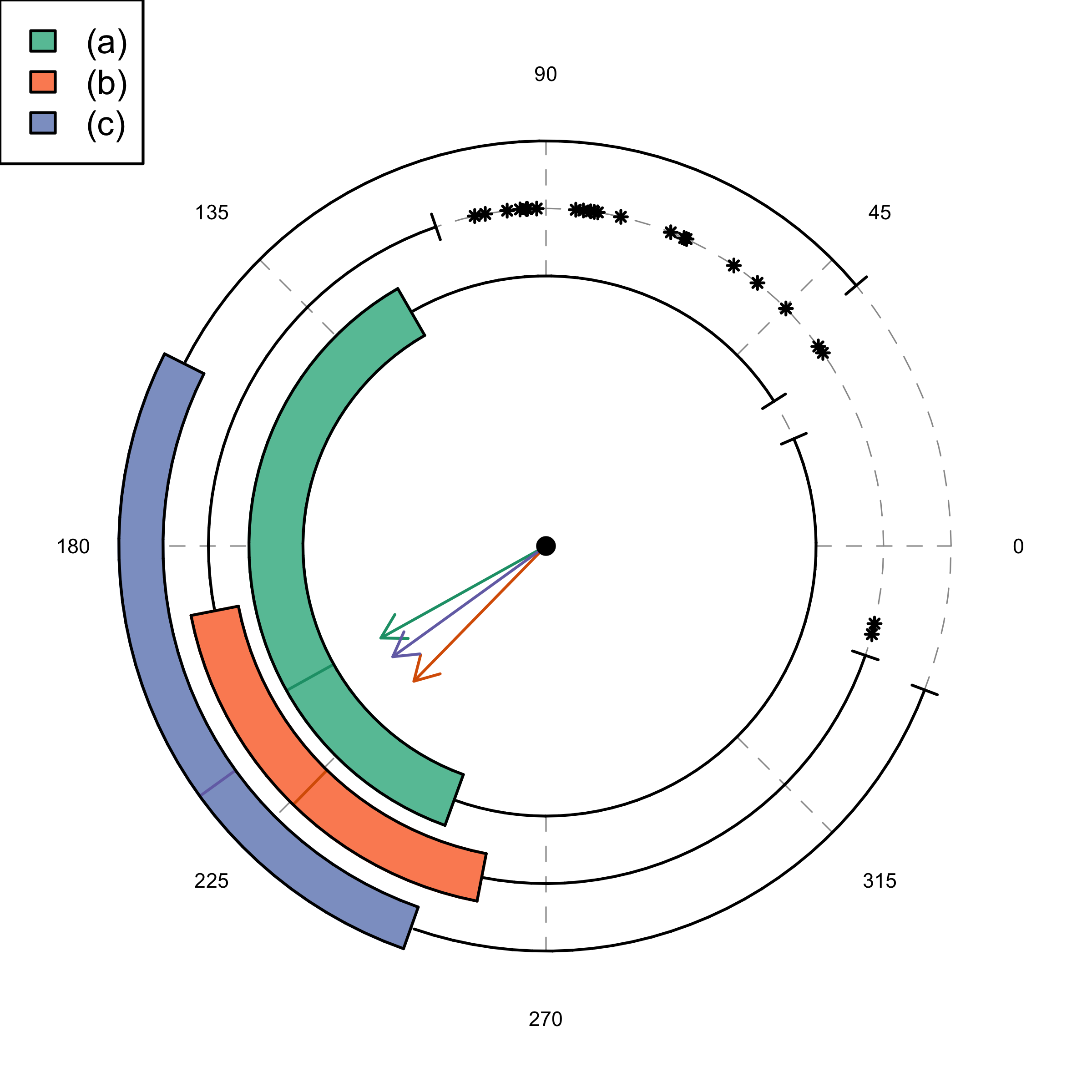}
  \caption{Grouped circular boxplot of the three-groups data in Figure~\ref{fig1}. We see that the medians and spreads of the three groups are easily compared (and distinct).
}
\label{fig3}
  \vspace{-0.1in}
\end{figure}

We revisit the three samples of Figure~\ref{fig1} and illustrate them in Figure~\ref{fig3}, which clearly shows the similarities and distinctions between the three
groups. For instance, we can readily see that all three medians are  distinct even though close, and that the first sample is skewed more than the third sample. Further, 
in Figure~\ref{fig3}, boxes drawn further from the centre of the circle are thinner than those inside the plot, providing a clearer perception.
This clarity in comparisons between the distribution of data across different groups is the main contribution here. We now discuss using quartile plots when we have a large numbers of circular boxplots.
\subsubsection{Quartile plots for circular data}
When we have many groups, it is not meaningful to scale the width of each box by its distance from the center. So, we adapt \citet{tufte1998}'s
quartile plot which replaces the idea of having boxes in a (linear) boxplot with a line of heavier weight over the middle half of the distribution.
Although the goal behind the quartile plot was to make the boxplot minimal and
to maximize ``data-ink,'' we can adopt the strategy here because it reduces the issues created by the width of the circular boxplots at different radii since an arc of minimal length does not have much by way of area. Then, we draw a single thick line in lieu of the box, with thinner grey lines as whiskers, producing a circular quartile plot as will be illustrated  in Section \ref{wind}.
\subsection{Illustrative examples}

\label{illustration}
We now illustrate the benefits of our development in effectively visualizing grouped angular data, and in  clarifying various features across different groups and conditions.

\subsubsection{Motor resonance data}
\label{motor}
\begin{figure}[h]
\centering  \includegraphics[width=0.65\linewidth]{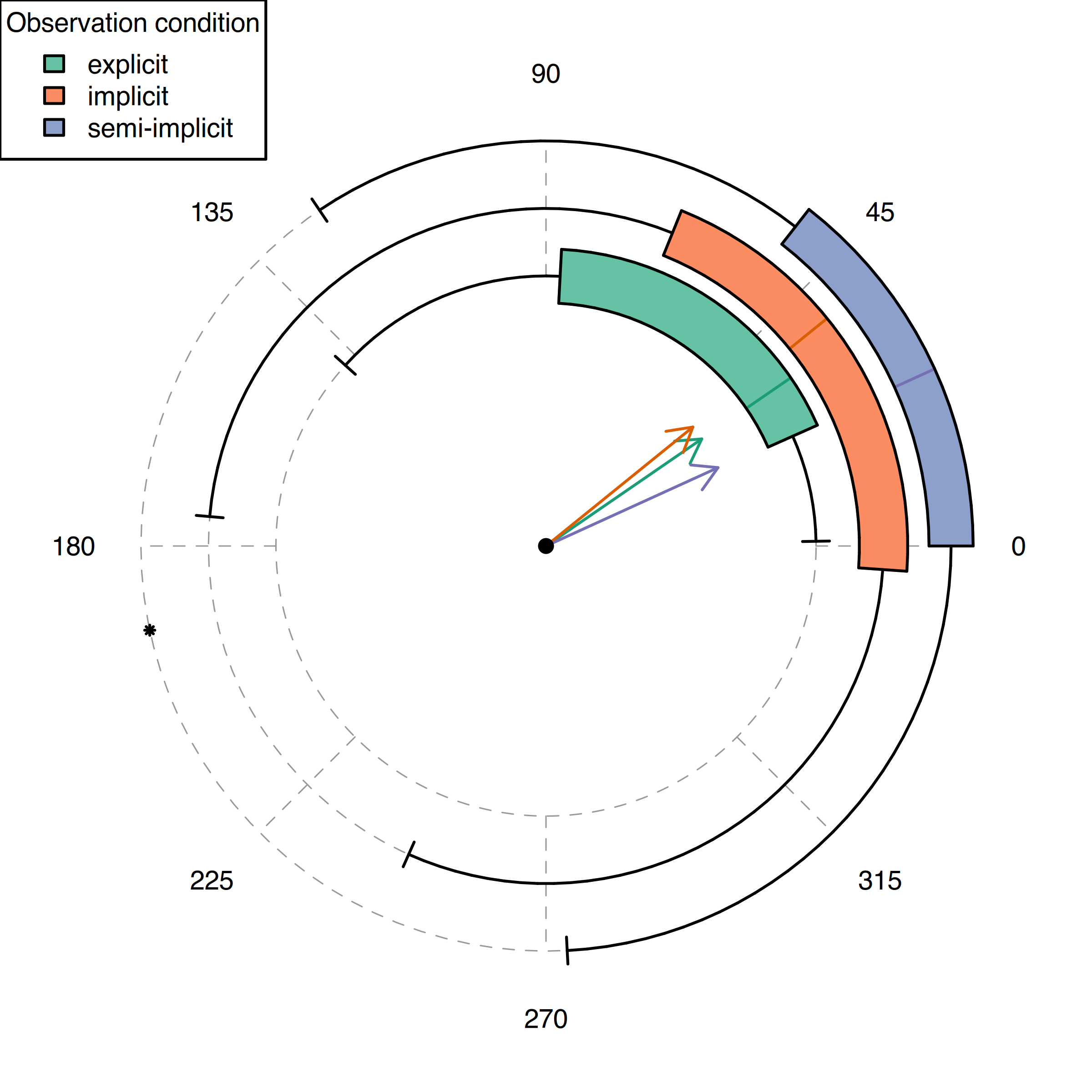}
\caption{ Grouped circular boxplot of the motor resonance data. We notice
that the implicit group is the most out-of-phase, with observations spanning a
large arc around the plot, followed by explicit, then semi-implicit. The
implicit group also has the most spread among groups. }
\label{fig4}
\end{figure}

We consider the  motor resonance data presented in \cite{puglisi2017}, where
an experiment was conducted by having subjects move their hands in a direction
subject to a given condition.
The first condition, called ``explicit observation,'' directed subjects to
observe the hand of another individual, called the mover, explicitly before
moving their own.
The second condition, ``semi-implicit observation,''
required subjects to observe the mover's hand implicitly.
The final condition required subjects to perform the task independent of the
mover and is called ``implicit observation.''
The observed data were recorded as a phase
difference between the hands of the mover and the observer, an angular metric
measured in degrees.
The experimenter hypothesized that hand movements between observers and movers
will have the strongest synchronization in the explicit case, followed by the
semi-implicit and implicit cases. Figure 3 of \citet{cremers2018} provides three separate side-by-side plots of the data for each group to compare them visually.
The difference between the explicit observation and the other two groups is clear in
their display, but the differences between the semi-implicit and implicit observation groups is more difficult to discern. Our grouped circular boxplot of Figure~\ref{fig4},
provides an easier comparison summary of the phase differences between groups. In particular, the key distinctions
between the semi-implicit and implicit groups are easily identified.
\subsubsection{The peak phases of circadian clock genes}
\label{gene}
\begin{figure}[!ht]
  \includegraphics[width=\linewidth]{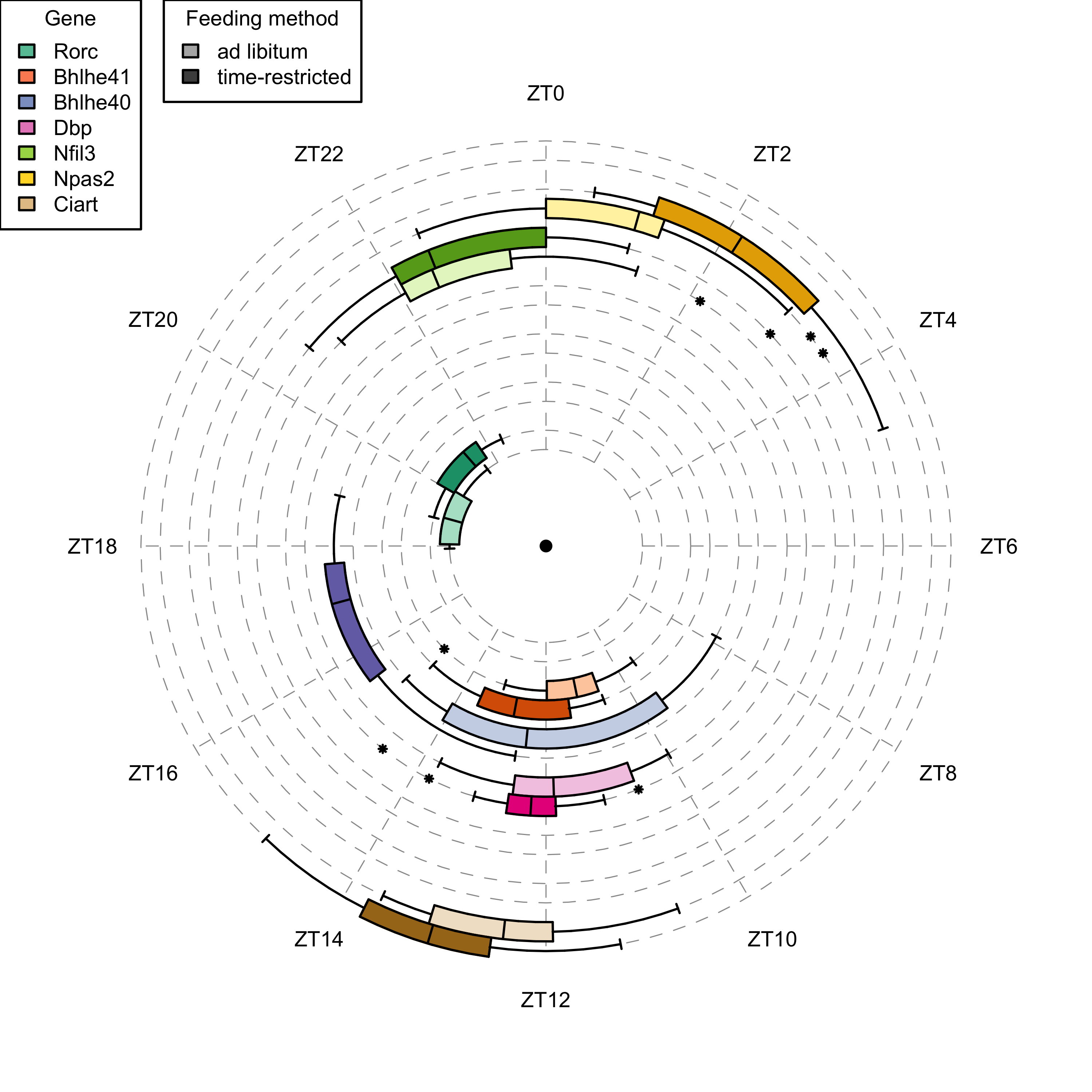}
\caption{ Distribution of peak phases for ancillary clock genes cycled in
{\it ad libitum} and time-restricted feeding of mice. }
\label{fig5}
\end{figure}

We next revisit a dataset containing peak phases of seven circadian clock
genes in tissues from mice subjected to either {\em ad libitum} or
time-restricted feeding \citep{deota2023}.
The data are provided in degrees corresponding to measurements
on a 24-hour {\em zeitgeber time} (ZT) clock cycle.
The authors present several grouped circular boxplots in their~\citep{deota2023} Figure 6.
However, note that a multiplicative factor of 1.5 is used to identify fences in
their display, generally used for linear data coming from a normal distribution,
as discussed previously.
Additionally, it is unclear if the unplotted outlying points are included in
their calculations or if the summary values assumed linear rather than circular statistics. 
Figure~\ref{fig5} combines the separate grouped circular boxplots from Figure 6 of
\cite{deota2023}, with boxes corresponding to the same genes grouped and
shaded based on feeding type.
This plot shows the differences in expressions between the two
feeding types more clearly than the side-by-side circular boxplots in
\citet{deota2023}.
\citet{deota2023} hypothesized that in most tissues, the
rhythms of gene expressions are not solely clock-dependent and that feeding
cycles play a role in regulating this rhythmicity.
This hypothesis can be supported by Figure~\ref{fig5}, where the peak
phase times from time-restricted feeding generally come later than those from
{\em ad libitum} feeding. Notably, the lighter-hued boxplots are, for the most
part, shifted clockwise to their dark counterparts.
Thus, replacing the obesogenic {\em ad libitum} feeding
cycle with a time-restricted cycle can help to regulate gene expression in
peripheral organs, potentially leading to positive health outcomes.

\subsubsection{\hspace{-0.1in} A deep dive into temporal wind direction distribution}
\label{wind}
Understanding the pattern of wind direction over time is often of interest for many reasons,
such as determining the direction in which a power-generating wind turbine should be pointed.
In this case, we may use discretized time values as a grouping variable.
When comparing the distribution of wind direction over time, wind rose diagrams
\citep{sundari2020, arteagalopez2021} for each time period (for example, months)
are employed and plotted side-by-side.
While these diagrams have their benefits, such as allowing for the inclusion of wind
speed in their display, they also suffer from problems due to their
histogram-like nature, for instance by bringing along the need for choice in bin width for
direction \citep{wells2000}.
On the other hand, if the interest is solely on the distribution of wind direction, a circular boxplot avoids a need for user-selected graphical parameters.
Naturally then, a grouped circular quartile plot allows for easy comparison of
direction across large spans of time, as seen in this application that displays
and interprets wind direction over the year  2023 in Ames, Iowa, USA and Jamshedpur, Jharkhand, India.

The data are as described in the Data availability statement section of this article.
For both stations and as is the norm, the stations do not report values with wind speed below two knots and consider such weather to be calm.
A wind direction of $0^\circ$ indicates no measurable wind, while a
direction of $360^\circ$ means wind from true north.
Because a reading of $0^\circ$ does not contain information on wind direction, we discard these observations in our analysis. There are also a few nonreports that are omitted.

\begin{figure*}[!ht]
  \vspace{-0.2in}
  \centering
  \subfloat[Ames wind direction]{\label{fig:2d-ames-wind}
\includegraphics[width=0.5\linewidth]{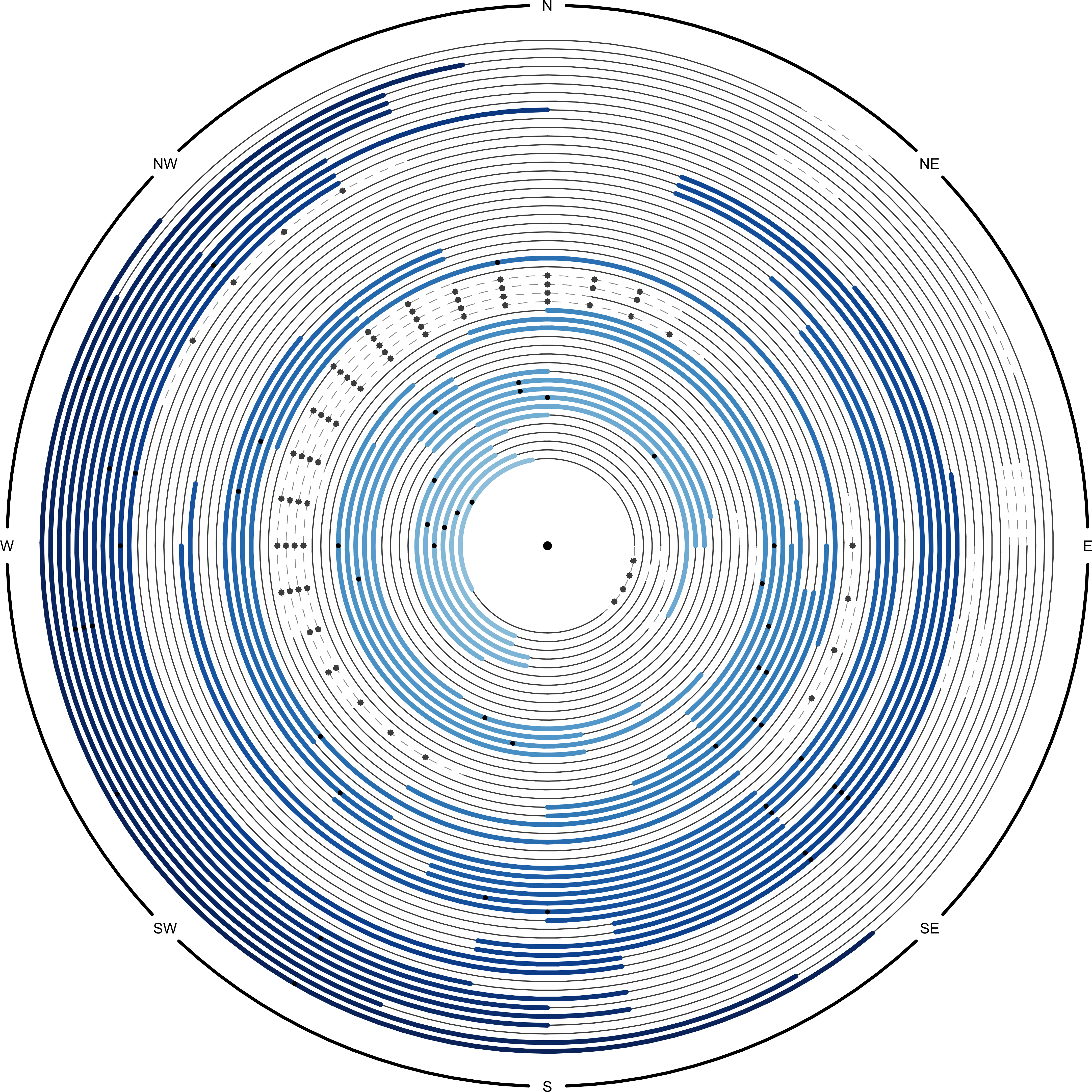}
}%
\subfloat[Jamshedpur wind direction]{\label{fig:2d-vejs-wind}
\includegraphics[width=0.5\linewidth]{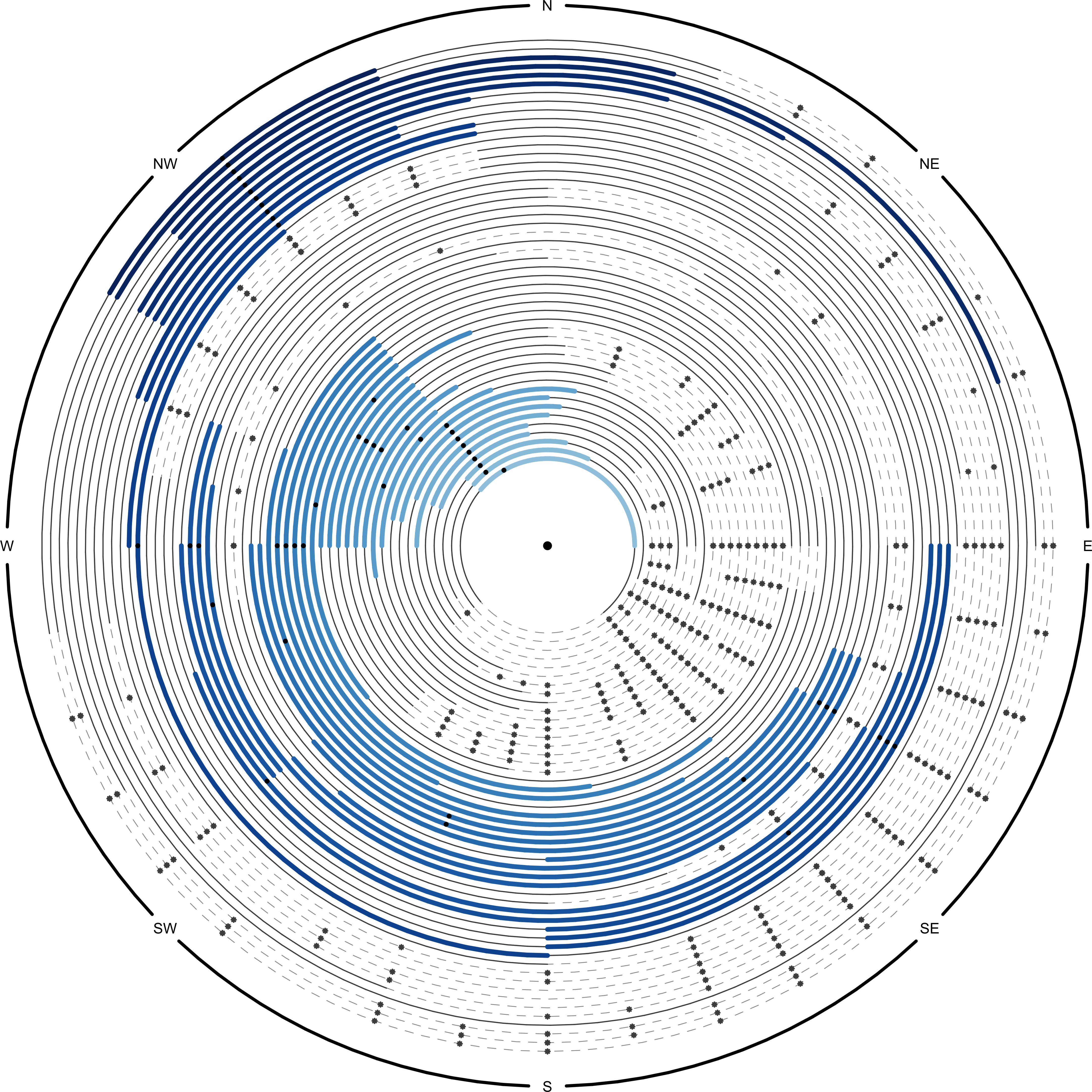}
}\\
\subfloat{
\includegraphics[width=\linewidth]{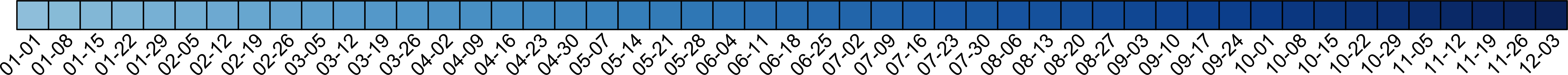}
}\\
\vspace{0.5em}
\caption{Distribution of wind direction over time in (a) Ames, Iowa and (b) Jamshedpur, India using grouped quartile plots. The legend indicates the starting day of the four-week period for each color.}
\label{fig:2d-wind}
\end{figure*}

Figures~\ref{fig:2d-ames-wind}
and \ref{fig:2d-vejs-wind} display the temporal nature of  wind direction in Ames and Jamshedpur over the course of 2023.
Here, each circular quartile plot represents a rolling four-week distribution (there are 
49 rolling wind direction distributions in all). 
Figure~\ref{fig:2d-ames-wind} displays  massive variability in wind direction during the year. The distribution largely spans the entire circle, as  seen from the span of the boxes and whiskers. Further, in the cold winter months of January, the wind direction is
mostly from the west but moves to the north as the year progresses.
Spring and summer have a lot of variability: indeed, there are times in the summer when the line representing the extent of the first and third
quartiles spans almost half of the circle.
As summer gives way to fall, the wind is more from the east and the southeast, but with fall, the direction changes back to the west.

In contrast, the four-week rolling distributions of wind direction in Jamshedpur, India are more concentrated by season, as indicated by the grouped circular quartile plots in Figure~\ref{fig:2d-vejs-wind}.
In the early part of 2023 (the cold season), the high pressure area is in the
Punjab in the northwest. The winds mostly flow from that
direction as the air is warmer in the Bay of Bengal (to the southeast of
Jamshedpur) and rises, and the cold air rushes in to take its place.
As the year rolls on, the hot season kicks in, and we can see the rise of the
``loo,'' or the hot and dry wind that blows from the plains of the northern
Indian subcontinent in May and June, and usually in the
afternoons~\cite{rana07}.
The heat eventually shows up as an intense area of low pressure in the Punjab,
causing the hot air to rise, with the cooler moisture-laden air from the Bay of
Bengal now rushing in to take its place.
Note that the southeast trade winds that are usually south of the equator move
up in the summer, so these winds get a further fillip by the above low-pressure
area.
This phenomenon, which essentially reverses the wind trajectories of the cold
season, gives rise to the southeast monsoon, and the transition is
quite clearly indicated in the plot.
These winds continue till around September after which the northwestern winds
again pick up as we go into the cold season. Thus, we see that our grouped circular quartile plots show interpretability and understanding of  wind direction  over the year.

\section{Displaying periodic angular data}
\label{sec:3d}
Figure~\ref{fig:2d-wind} shows angular distributions over time that has a periodic component. While the data and boxplots themselves indicate the periodicity of weather by year, the display itself does not make it easy to assess the cyclical nature of wind direction over the weeks of the year. We address this shortcoming by developing a toroidal display (in 3D) to also incorporate a periodic component (here, for time). Specifically, while the circular boxplots for each four-week rolling distribution can be placed in the torus' {\em toroidal} angle, the axis of each of these boxplots can be placed along its {\em poloidal} angle. We describe our methods and algorithm next. 
\subsection{Toroidal boxplots and quartile plots}
\label{3d-plot}
\begin{figure}[h]
   \centering
  \vspace{-0.1in}
  \includegraphics[width=0.65\linewidth]{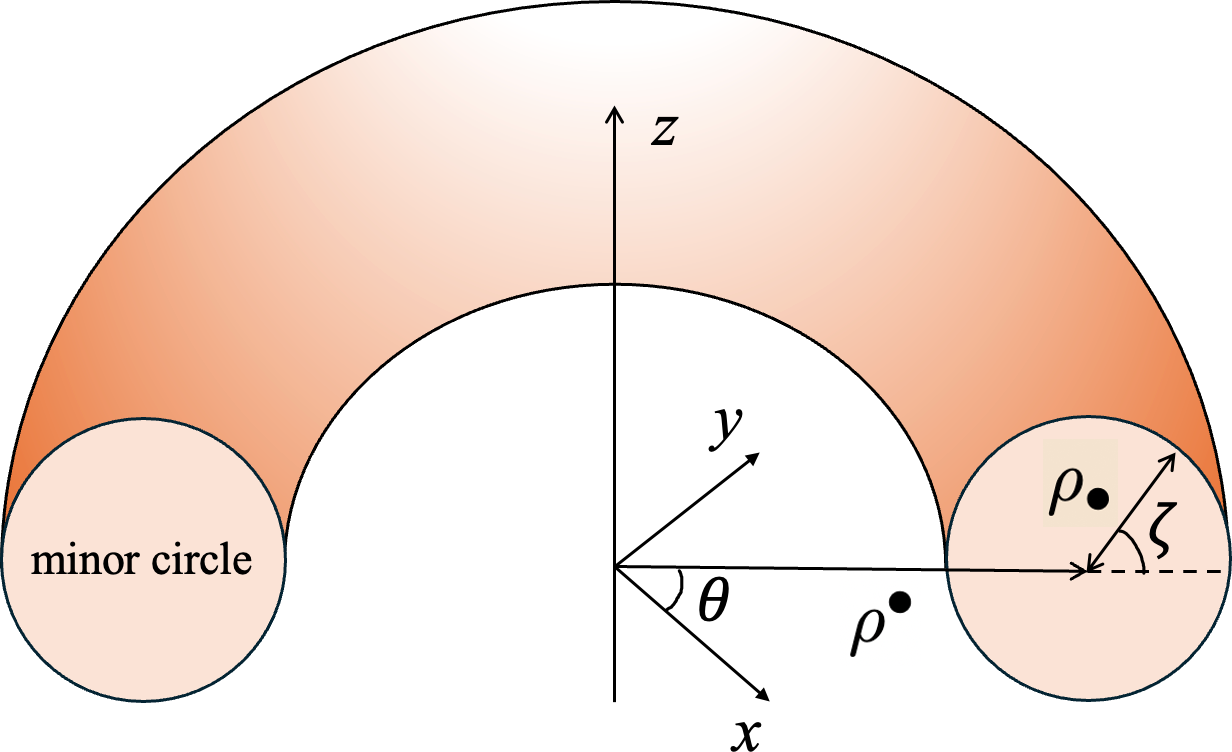}
\caption{The toroidal coordinate system.}
\label{torus-fig}
\end{figure}

We use a toroidal coordinate system \cite{rose2017}  for our correspondingly-named toroidal boxplots and quartile plots. (Our mechanism for constructing a boxplot and quartile plot is essentially the same, so we describe construction only through that of a toroidal boxplot.) In geometry, a ring torus is generated by revolving a circle of radius $\rho_\bullet$ in 3D space around an axis with radius $\rho^\bullet$ without intersection.  This framework defines a toroidal coordinate system with major and minor radii $\rho^\bullet >0$ and $\rho_\bullet\in(0,\rho^\bullet)$, and the poloidal and toroidal angles $\zeta\in(0,2\pi]$ and $\theta\in(0,2\pi]$ as shown in Figure~\ref{torus-fig}.
Accordingly, the transformation from the toroidal coordinates to the standard Cartesian coordinates $(x,y,z)$ is given by,
\begin{equation}\label{torus-coords}
    \begin{split}
        x &= (\rho^\bullet + \rho_\bullet\cos{\zeta})\cos{\theta}\\
        y &= (\rho^\bullet + \rho_\bullet\cos{\zeta})\sin{\theta}\\
        z &= \rho_\bullet\sin{\zeta}
    \end{split}
\end{equation}
For circular data from $m$ groups (where the groups have a periodic component), we use the toroidal angles to represent the circular data in each group, and the poloidal angle to denote  group periodicity. Operationally, we divide the minor circle of the ring torus into $m$ equal parts with points of division $(x_i,y_i,z_i),i=1,2,\ldots,m$, then, we determine the distance of the $i$th circular boxplot from its center by $\sqrt{x_i^2+y_i^2}$
and the position of the plot in 3D space by $z_i$. Consequently, we obtain grouped circular boxplots distributed uniformly around a torus surface. Algorithm~\ref{alg-torus-mesh} summarizes the construction of grouped toroidal boxplots.
\begin{algorithm}[!ht]
\caption{Toroidal boxplots for $m$ data groups}\label{alg-torus-mesh}
\begin{algorithmic}[1]
\Require $\rho^\bullet>0$, $0<\rho_\bullet<\rho^\bullet$, $m \geq 3$
\State Compute $\theta_1,\theta_2,\ldots,\theta_m$ and $\zeta_1,\zeta_2,\ldots,\zeta_m$ as $m$ points of division of length $2\pi$.
\State Compute $(x_i,y_i,z_i),i=1,2,\ldots,m$ using the coordinate transformation in \eqref{torus-coords}.
\State Compute the radius of the $i$th boxplot as $\rho_i = \sqrt{x_i^2+y_i^2}$.
\State Given $\rho_i$, draw the $i$th grouped circular or quartile boxplot as described in Section \ref{group-circ-boxplot}.
\State Place the $i$th boxplot obtained from Step 4 on the 3D torus surface according to coordinate $z_i$.
\end{algorithmic}
\end{algorithm}

\subsection{Illustrative examples}
\begin{figure*}[!ht]
\centering
\subfloat[Ames wind direction]{\label{fig:3dames}
\animategraphics[controls={play,step},loop,width=0.5\linewidth]{1}{figs_png/ames-wind/ames-wind-3d-}{1}{53}
}%
\subfloat[Jamshedpur wind direction]{\label{fig:3dvejs}
\animategraphics[controls={play,step},loop,width=0.5\linewidth]{1}{figs_png/vejs-wind/vejs-wind-3d-}{1}{53}
}\\
\subfloat{
\includegraphics[width=\linewidth]{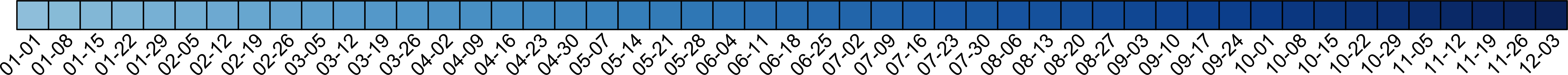}
}
  \vspace{1em}
\caption{Visuanimations of the distribution of wind direction over 2023 in (a) Ames, and (b) Jamshedpur using toroidal quartile plots. Click on the central button of each figure to play the animation.}
\end{figure*}

We revisit the datasets of Figure~\ref{fig:2d-wind}
using toroidal quartile boxplots. 
Figures~\ref{fig:3dames} and 
\ref{fig:3dvejs}
provide corresponding toroidal quartile plot displays to Figures~\ref{fig:2d-ames-wind} and \ref{fig:2d-vejs-wind}. 
These figures are visuanimations of 3D displays and show the toroid from several pre-set angles, but we also point the reader to 
an online HTML resource at \url{https://fanstats.github.io/CircularDataViz.github.io/} that allows for a fuller experience of the benefits of 3D visualization. 

Both displays show the benefits of incorporating the periodicity of time in the display, as we see that the cyclic nature of wind direction over the course of the year is clearly displayed. 
To see this, note that the darkest colors of the quartile plots are when the year begins while the lightest shades are towards the end of the year. Both our toroidal quartile plots show these boxplots to align close to each other, therefore capturing the periodicity of wind direction over the course of the year. Our toroidal quartile plots therefore display both the angular dimension of the wind distribution, and the periodicity of the distribution over the course of the year, and provides a powerful tool for more fully  displaying these two aspects of these datasets.

\section{Discussion}
\label{disc}
This article discusses methodology for displaying the distribution of grouped and periodic  circular data in a way that captures the circularity of such datasets. 
We demonstrate that \citet{buttarazzi2018}'s circular boxplot for one population can be extended for  multiple groups, which may themselves have periodic components. 
In so doing we address the problem of identical boxplots appearing visually
different based on their distance from the center of the plot, as shown in Figure~\ref{fig2}.
We demonstrate the value of displaying concentric circular
boxplots rather than side-by-side plots in Figures~\ref{fig4} and~\ref{fig5} for distinguishing angular distributions from different groups. 
The use of concentric boxplots however brings challenges of perception that we address through ensuring that widths of the boxes are inversely proportional to the square root of their distance from the center. 
Our visualization is validated by the results of a survey done to evaluate how an user will perceive the distribution from these boxplots. We also develop circular quartile plots for when 
the number of groups is too large to be meaningfully displayed by way of boxplots. Finally, we develop 3D toroidal boxplots and quartile plots to account for the periodicity of distribution of directions over time. Our R~\citep{R} package {\bf CircularBoxplots} implements our methods and is used to illustrate our methodology on four datasets. Our displays clearly provide improved understanding and interpretability of grouped and temporal angular datasets. 

There are a few areas that could benefit from further attention. Our methods have been in the context of angular data. However, it would be worthwhile to see if we could also provide 
displays for data from spheres, or for datasets that have both linear and circular measurements.

\section*{Acknowledgments}
The research of the  third author was supported in part by the United States Department of Agriculture (USDA) National Institute of Food and Agriculture (NIFA) Hatch project IOW03717. The content of this paper is however solely the responsibility of the authors and does not represent the official views of the USDA.

\section*{Supplementary Materials}
The following supplementary materials are available and contain:
\begin{enumerate}
  \item A compressed file archive provided in (\texttt{code-for-circular-boxplots.tgz}) that contains all the code required to reproduce all the results and figures for each section in this article.

  \item \textbf{CircularBoxplots}: A R package implementing the display methods
    introduced and discussed in this article, with developmental version  available at
    \url{https://github.com/jdberlinski/CircularBoxplots}.
\end{enumerate}

\section*{Conflict of Interest}
None.

\section*{Data availability statement}
The data used in this article are all publicly available.
The motor resonance dataset is available in the R package {\bf bpnreg} under the name Motor.
The circadian clock genes dataset is from \citet{deota2023} and also available as part of our package. 
The wind direction datasets are from Iowa State University's Iowa Environmental
Mesonet site at \url{https://mesonet.agron.iastate.edu/request/download.phtml}.
The Ames dataset is from the Automated Surface Observing System (ASOS), the
United States' flagship automated weather observing network.
Because it is an ASOS station, the location is at Ames Municipal Airport (ICAO,
or International Civil Aviation Organization, code AMW; latitude
$41.99263^\circ$N and longitude $93.62184^\circ$W),
with its primary function being collection of minute-by-minute weather data.
The Jamshedpur dataset was collected by the India ASOS network (or the similar
Automated Weather Observing System or AWOS) network.
The location for these observations is Sonari Airport in Jharkhand, India (ICAO
code VEJS) at latitude $22.81315^\circ$N and longitude $86.16885^\circ$E,
where data are collected half-hourly. The wind datasets are also available as part of our package.

\bibliography{ref}
\bibliographystyle{mytran}
\end{document}